\documentstyle[amssymb,preprint,aps]{revtex}

\begin{document}
\title{{\protect\Large RECIPROCAL SCHR\"{O}DINGER EQUATION: DURATIONS OF
DELAY AND OF FINAL STATES FORMATION IN PROCESSES OF SCATTERING}}
\author{\bf Mark E. Perel'man }
\address{\it Racah Institute of Physics, Hebrew University, Jerusalem, Israel}
\date{\today}
\maketitle

\begin{abstract}

The reciprocal Schr\"{o}dinger equation $\partial S(\omega ,{\bf r}%
)/i\partial \omega =\widehat{\tau }(\omega ,{\bf r})\ S(\omega
,{\bf r})$ for $S$-matrix with temporal operator instead the
Hamiltonian is established via the Legendre transformation of
classical action function. Corresponding temporal functions are
expressed via propagators of interacting fields. Their real parts
$\tau _{1}$are equivalent to the Wigner-Smith delay durations at
process of scattering and imaginary parts $\tau _{2}$ express the
duration of final states formation (dressing). As an apparent
example,
they can be clearly interpreted in the oscillator model via polarization ($%
\tau _{1}$) and conductivity ($\tau _{2}$) of medium. The $\tau
$-functions are interconnected by the dispersion relations of
Kramers-Kr\"{o}nig type. From them follows, in particular, that
$\tau _{2}$ is twice bigger than the uncertainty value and thereby
is measurable; it must be negative at some tunnel transitions and
thus can explain the observed superluminal transfer of excitations
at near field intervals (M.E.Perel'man. In: arXiv. physics,
Gen.Phys/0309123). The covariant generalizations of reciprocal
equation
clarifies the adiabatic hypothesis of scattering theory as the requirement: $%
\tau _{2}\rightarrow 0$ at infinity future and elucidate the
physical sense of some renormalization procedures.

\end{abstract}


\section{INTRODUCTION}

In the classical physics are used different ''spaces'' or sets of variables,
best suited to consideration of different problems. For these aims are
formulated the methods of transition between corresponding forms of
dynamical equations, performing by Legendre transformations of action
functions, and corresponding transformations in thermodynamics. In the
quantum theory for similar aims are using expansions over complete sets of
suitable functions: exponents, i.e. the Fourier transformation from $(t,{\bf %
r})$ to $(E,{\bf p})$ variables and vice verse, the Legendre functions ,
i.e. transition to moment variables, etc.

However, as far as we know, in quantum theory the possibilities of direct
Legendre transformations, which must lead to reciprocal equations, are not
considered. It can be partly attributable to the distinction with classic
fields, where all equations are directly expressed via action functions, but
equations of quantum dynamics are expressed via exponents of these
fundamental magnitudes, which causes some complications at deducing the
reciprocal expressions.

Are needed such reciprocal expressions or not, can they lead to some new
physical results, or they will represent only a methodological interest? It
is a crucial question and we shall demonstrate that the reciprocal
Schr\"{o}dinger equation permits, at least, to understand and investigate
the duration of interactions. Such approach does not exclude, of course,
possibility of performing these investigation by more common methods, but
shows some prospects of developing theory.

The consideration of several problems of durations of interaction have
aroused a lot of research and their discussions are continuing. For
introduction into these problems we give overview of the status of some
duration presentations [1-17]. As will be shown, the joined consideration of
two magnitudes, the duration of delay at scattering processes and the
duration of final states formation, their ''dressing'', essentially simplify
problems. These two magnitudes, usually examined separately, are unified by
the reciprocal Schr\"{o}dinger equation and just this circumstance clarifies
the necessity of proposed Legendre-type transformation consideration.

The notions of delay (waiting time) in the process of signal transfer and of
duration, needed for the final form restoration, were naturally perceived in
classical physics, e.g. for oscillating systems; thus they were
experimentally and theoretically investigated in the electrical engineering
and in the acoustics, e.g. [1]. But in quantum theory during long time there
was actually implied that all problems, connected with particles and states
formation and with coupling of particles during scattering processes, should
be restricted only and only by the frameworks of uncertainty principles.

The first (semi-qualitative) consideration of time delay in processes of
tunneling had been performed, as far as we know, by Bohm [2]. The more
constructive and physically more transparent notion of time delay under the
elastic scattering was introduced by Wigner [3] through the derivative of
partial phase shifts: $\tau _{l}(\omega )=d\delta _{l}/d\omega $. This
expression was generalized by Smith [4] via $S$-matrix of scattering:

$\tau $ $=$ Re $\frac{\partial }{i\partial \omega }\ln S$. \qquad \qquad
(1.1)

On the base of (1.1) Goldberger and Watson derived a ''coarse-grain''
Schr\"{o}dinger equation [5], it shown the generality of this magnitude. But
at their approach the magnitude (1.1) had been introduced artificially, by
decomposition of the logarithm of Fourier transformed response function $%
R(t) $ of linear relation

$O(t)=R(t)\otimes I(t)\equiv \int dt\prime R(t-t\prime )\ I(t\prime )$
\qquad \qquad (1.2)

near the selected frequency, without discussion of its imaginary part,
higher terms and dependence on space variables.

In the case of photonic processes it is intuitively evident, that the delay
time is the duration between absorption of single photon and its reemission
or vice verse by a bound or free electron (this time duration coincidences
with (1.1) and will be denoted as $\tau _{1}$). It should be underlined that
this time duration could be deduced in the course of QED calculations. So
after summarizing of the complete sequence of the main $S$-matrix terms for
some multiphoton processes the evaluated infinite series may be reformulated
via the parameter $j/j_{0}=j\sigma _{tot}\tau _{1}$ with $\tau _{1}$, the
flux density $j$ and the total cross-section $\sigma _{tot}$ of single $%
\gamma $-$e$ scattering, and this naturally appearing parameter determines
the thresholds of some multiphoton processes saturation and of the new
channels opening [6, 7]. It demonstrates that the duration time expressions
are implicitly contained in the QED and therefore the corresponding
magnitudes should be recognized in the common theory.

Another approach, which seems at first glance distinctive from the
Wigner-Smith one, was suggested by Baz' [8] for the consideration of
nonrelativistic tunneling processes. This approach consists in attributing
to the scattering particle some moment, e.g. magnetic, and in analyzing its
turns at the scattering process (the method of ''Larmor clocks''), some its
variants are reviewed, in particular, in [9] and in several articles in
[10]. This method will be very shortly discussed later with demonstration of
its principal identity with the Wigner-Smith approach.

But even earlier Frank had been forced to introduce in the theory of
\^{C}erenkov radiation the notion of path length (or duration), necessary
for the extended formation of real photon by ''superluminal electron in
media'' [11]. Without such concept was completely incomprehensible the
discrete character of this emission, and Frank had been forced to consider
the interference picture of continuously emitted (virtual) waves, which
leads to the real emission at resonance conditions.

Then, independently, Ter-Mikaelyan [12] and Landau and Pomeranchuk [13] had
considered the duration of photon formation in the theory of bremsstrahlung:
it is the time duration needed for a virtual coat formation around particle,
its dressing (cf. the reviews [14]). But as must be underlined, the concept
of particle or state formation has more general sense: so Moshinsky had
calculated, through the non-stationary Schr\"{o}dinger equation, the
duration necessary for establishment of the certain state of electron after
its transition onto the upper level [15], and it is also the duration of
final state formation.

Notice that the time duration of final state formation can be, in principle,
naturally measurable in the processes of multiphoton ionization [7, 16]: it
must be such extended period of time, during which the photoelectron,
already absorbing enough energy for liberation, is yet in the virtual state,
retains its association with the atom/ion till gain for the moment
corresponding to the absorbed energy. In such virtual state electron can
absorb additional, above threshold photons. Thus the multiphoton processes,
except some cases of high harmonics generations, require the correlation of
two independent, generally speaking, time durations: the duration of energy
absorption depending on the photon flux density and so on, and the duration
of corresponding moment accumulation, depending on interaction with
surrounded particles and fields.

Therefore multiphoton processes seem exceptionally interesting for all
quantum theory: in these processes are examined the concept of virtual coats
of particles and the dynamics of their formation.

All results cited above, which were established in various and, we think,
artificial methods, can be calculated at the unified and simplified way by
the expression

$\tau _{2}$ $=$ Im$\frac{\partial }{i\partial \omega }\ln S$ \qquad \qquad
(1.3)

through the response function or matrix element of transition [15].

This expression may be formally considered as the analytically complementary
to the Smith's formulae (1.1). Since (1.3) can be rewritten as $\tau _{2}$ $%
= $ $\partial \ln \left| S\right| /\partial \omega )$, it can be considered
as the measure of incompleteness of the final state or of the outgoing
dressed particle formation. As far as we know, the similar expression for $%
\tau _{2}$ was introduced, for the first time, by Pollak and Miller [17] and
was interpreted as the duration of tunneling process.

The main purpose of the paper consists in some simultaneous refinement of
both notions, of delay and of formation of final state, and in the revealing
of their place in the common scattering and general field theories. Thereby
the investigation of several approaches to revealing a latent, as though,
existence of temporal expressions or their equivalents in common theories is
needed.

The most natural way in this direction can begin, as represented, with the
more general formulation of considered magnitudes. So both definitions (1.1)
and (1.3) can be formally combined as $\tau (\omega ,{\bf r})\equiv \tau
_{1}+i\tau _{2}=\frac{\partial }{i\partial \omega }\ln S(\omega ,{\bf r})$.
Then it must be attempted to consider this expression as the consequence of
some equation,

$\frac{\partial }{i\partial \omega }S(\omega ,{\bf r})$ = $\widehat{\tau }%
(\omega ,{\bf r})\ S(\omega ,{\bf r})$. \qquad \qquad (1.4)

Formally this relation seems analogical to the Schr\"{o}dinger equation for $%
S$-matrix, but rewritten via some transformation of $x_{\mu
}\longleftrightarrow p_{\mu }$ type and with a some ''temporal'' operator $%
\widehat{\tau }(\omega ,{\bf r})$ instead the Hamiltonian.

The proposed transition to new variables can be performed by the Legendre
transformation in classical theory, at least. On the other hand it can be
performed by the Fourier transformation of response function in (1.2), and
in the Section 2 both approaches are examined: they lead to approximately
identical results.

The main properties of unified temporal functions $\tau (\omega ,{\bf r})$,
their simplest interpretation and their interrelations with the uncertainty
magnitudes are considered in the Section 3. As these functions are causal,
for them can be established the certain dispersion relations and
corresponding sum rules (Section 4), that demonstrate some principal
properties of temporal functions. The received results are discussed in the
Section 5 on the example of the simplest oscillator model of medium; it
descriptively reveals the physical sense of both temporal functions.

As the temporal parameters can be considered as the results of interference
of waves, coming from different points, it seems that the suiting functions
for their comparative investigation should be the Wigner functions (Section
6). Their consideration shows that the expressions of temporal functions are
close to propagators (resolvents or Green functions), and it will be proven
in the Section 7 in the scope of formal theory of scattering.

In the Section 8 temporal functions and their covariant forms will be
considered by the methods of quantum field theory, and it will be proven
that the equations of (1.4)-type can be generalized till completely
covariant analogues of the Tomonaga - Schwinger equations. Therefore it will
be shown that the developed theory can be considered as the justification
for the adiabatic hypothesis of quantum theory of interactions and as its
generalization; it permits to understand the physical sense of such formal,
as usually seems, mathematical procedure.

The Section 9 is devoted to some problems of QED. Their considerations are
continued in the Section 10 by interpretation of renormalization procedures,
the Pauli-Villars and the subtraction methods, and, more generally, the
renormalization group equations via the temporal functions.

In the Conclusions are summed up the main results and are contemplated some
perspectives of further investigations.

\qquad

{\large 2. LEGENDRE TRANSFORMATIONS AND FOURIER TRANSFORMATION}

The basic equation of quantum dynamics for the evolution operator ($S$%
-matrix),

$i\frac{\partial }{\partial t}S={\bf H\ }S$, \qquad \qquad (2.1)

can be formally deduced from the Hamilton-Jacobi equation for classical
action function and Hamiltonian,

$(\partial /\partial t)S_{cl}(q_{i};\partial S_{cl}/\partial q_{i};t)={\bf H}%
_{cl}(q_{i};\partial S_{cl}/\partial q_{i};t)$.\qquad \qquad\ (2.2)

For such transition is used the Schr\"{o}dinger-type heuristic substitution:

$S_{cl}\rightarrow \ i\hbar \ln \{S(t,r)/\hbar \}$ \qquad \qquad (2.3)

with the determination: $d\ln S\equiv (dS)\ S^{-1}$ (below $c=\hbar =1$) and
classical variables $x$, $p$ are replaced by corresponding operators.

The transition to new variables in the classical action function is realized
by the Legendre transformation (e.g. [18]):

$S_{cl}(q\prime ;p\prime ;t\prime )=S_{cl}(q;p;t)-\sum (q\prime q+p\prime
p+t\prime t)$. \qquad \qquad (2.4)

Thus the canonical transformation from the time variable $t$ to the variable
of energy, $t\rightarrow \ t\prime \ ={\bf H}\rightarrow \ E$, in the
equation (2.2) results in

$S_{cl}(t;...)-{\bf H}t=S_{cl}^{L}(E;...)$, \qquad \qquad (2.5)

and the canonical equation (2.2) is transformed into the temporal
Hamilton-Jacobi equation:

$(\partial /\partial E)S_{cl}^{L}(E;...)=-T_{cl}(E;...)$, \qquad \qquad (2.6)

in which the role of Hamiltonian plays (classical) function of duration of
considered process. It leads to classical temporal Hamilton equations and so
on.

The employment of the analog of Schr\"{o}dinger-type substitution (2.3),

$S_{cl}^{L}(E;...)\rightarrow \ i\ln S^{L}(E,{\bf r})$, \qquad \qquad (2.7)

leads to the quantum equation (we use more familiar for such notations
symbol $\omega $ instead $E$):

$\frac{\partial }{i\partial \omega }S^{L}(\omega ,{\bf r})=\tau ^{L}(\omega ,%
{\bf r})\ S^{L}(\omega ,{\bf r})$, \qquad \qquad (2.6')

from which follows the determination of temporal function in accordance with
the Legendre transformation:

$\tau ^{L}(\omega ,{\bf r})$ $=\frac{\partial }{i\partial \omega }\ln
S^{L}(\omega ,{\bf r})$. \qquad \qquad (2.8)

The Legendre transformation can be performed at nonzero values of the
Hessian, i.e. the determinant consisted from second derivatives:

$J(\tau \rightarrow \omega )=(\partial _{t}\partial _{t}\ln S^{L})(\partial
_{{\bf r}}\partial _{{\bf r}}\ln S^{L})-(\partial _{{\bf r}}\partial _{t}\ln
S^{L})^{2}\neq 0$. \qquad \qquad (2.9)

It can be rewritten as

$J(\tau \rightarrow \omega )=\partial _{t}H\ \partial _{{\bf r}}{\bf P}%
-(\triangledown H)^{2}\neq 0\qquad \qquad $ (2.9')

and evidently determines conditions needed for possibility of introduction
of the temporal functions $\tau ^{L}(\omega ,{\bf r})$. Note that as the
Legendre transformation $L$ is performed by the involution operator, $%
L^{2}=1 $, this transformation does not change the magnitudes of observables
and Poisson brackets (commutation relations). Notice that the variation of
function $\tau ^{L}(\omega ,{\bf r})$ immediately leads to the Fermat
principle.

Further Legendre transformation ${\bf r}\rightarrow \ {\bf r}\prime \ ={\bf k%
}$ of the function $S^{L}(\omega ,{\bf r})$ leads to the equation:

${\bf \rho }^{L}(\omega ,{\bf k})=\frac{i\partial }{\partial {\bf k}}\ln
S^{L}(\omega ,{\bf k}),\qquad \qquad $(2.10)

which must correspond to an extent of interaction region in dependence on
energy-moment.

If the Legendre transformation (2.5) is considered as the sum of
infinitesimal canonical transformations, the expression (2.5) transfers,
evidently, into equation

$d\ S_{cl}(t;...)-d\ \sum_{k}\Delta _{k}({\bf H}t)=d\ S_{cl}^{L}(\omega
;...) $. \qquad \qquad (2.5'')

After passage to the limit, substitution of (2.7) and integration, this
expression leads to the representation:

$S^{L}(\omega ,{\bf r})=S_{0}(\omega _{0},{\bf r})\exp [-i%
\displaystyle\int %
_{\omega _{0}}^{\omega }\tau ^{L}(\omega ,{\bf r})\ d\omega ]$, \qquad
\qquad (2.11)

which can be considered as the general solution of the equation (2.6').
Integration in it goes from some $\omega _{0}$, at which $\tau ^{L}(\omega
_{0},{\bf r})$ is known, till the examined value $\omega $. Corresponding
general solution of (2.10) will be expressed via density of the 4-volume of
interaction.

Notice that the performed manipulations can be formulated as the
prescription: the Legendre transformation to new variables must be executed
in the exponents of quantum expressions.

Operators $(\tau ^{L},\ {\bf \rho }^{L})=(\partial /i\partial \omega ,\
i\partial /\partial {\bf k})$ form the 4-vector $\widehat{x}_{\mu }$,
corresponding to the equation:

$[(\tau ^{L})^{2}-({\bf \rho }^{L})^{2}-s^{2}]\ S^{L}(\omega ,{\bf r}%
)=-[\square _{\omega {\bf k}}+s^{2}]\ S^{L}(\omega ,{\bf r})=0$ \qquad
\qquad (2.12)

with a 4-interval $s$. It can be considered as the reciprocal one to the
Klein-Gordon equation and as the differential analogue, at $s^{2}\geq 0$, of
the relativistic generalization of Kramers-Kr\"{o}nig dispersion relations
[19, 20].

Let's consider now the function $S^{F}(\omega ,{\bf r})$, the Fourier
transform of response function of the linear relation for signal passed
through a uniform passive linear medium:

$O(t,{\bf r})=\int dt\prime \ d{\bf r}\prime \ R(t-t\prime ;{\bf r}-{\bf r}%
\prime )\ I(t\prime ,{\bf r}\prime )$. \qquad \qquad (2.13)

The law of energy conservation in classical theory or the unitarity
principle in quantum theory are responsible for the existence of partial or
complete Fourier transformations of this relation:

$O(\omega ,{\bf r})=R(\omega ,{\bf r})\ I(\omega ,{\bf r})$.\qquad \qquad
(2.13')

The logarithm of response function can be expanded near characteristic
frequency $\omega _{0}$ and this series can be restricted in some cases by
the first terms [4]:

$\ln R(\omega ,{\bf r})=\ln R(\omega _{0},{\bf r})+i(\omega -\omega _{0})\
\tau ^{F}(\omega ,{\bf r})+...$\qquad \qquad (2.14)

with the designation:

$\tau ^{F}(\omega ,{\bf r})=-i\partial _{\omega }\ln R(\omega ,{\bf r})$.
\qquad \qquad (2.15)

The subsequent terms of this series are expressed via derivatives of (2.15):
the first of them shows a temporal spread of signal and is considered in the
next Section.

Hence the response function can be expressed (approximately!) as

$R(\omega ,{\bf r})\simeq R(\omega _{0},{\bf r})\ \exp \{i(\omega -\omega
_{0})\ \tau ^{F}(\omega ,{\bf r})\}$.\qquad \qquad\ (2.16)

The comparison of (2.11) to (2.16) shows that the Fourier form of temporal
functions correspond to an averaged Legendre-type temporal functions and,
generally speaking, there are necessity for estimations of omitted terms of
(2.14) in comparison to (2.11).

In the classical theory of (2.13) temporal quantities can be evidently
deduced by direct expansion of responce function in (2.13') near the
separated frequency. But more interesting seems the revealing of temporal
magnitudes in another way. Let us consider the duration of rotation in the
classical mechanics, which is determined as

$T=2\int_{a}^{b}dx/v(x)=2m\int_{a}^{b}dx\ [2m\ (E-V(x))]^{-%
{\frac12}%
}$, \qquad \qquad (2.17)

where $v$ is the velocity of rotating particle, $E$ and $V$ are the complete
and potential energies. Via the action function $A=2\int_{a}^{b}p(x)dx$ the
duration of process is determined as

$T=4\partial A/\partial E$. \qquad \qquad (2.18)

The duration, for example, of the packet spreading over a system of
equidistant levels was determined in the ''old'' quantum mechanics as [21]

$\Delta T\ \symbol{126}\ 1/(\partial \Delta E/\partial A)\approx \partial
^{2}A/\partial E^{2}$.\qquad \qquad (2.18')

At transition from classical mechanics to quantum one in accordance (2.7) $%
A\rightarrow -i\hbar \ln (S/\hbar )$, and just this substitution leads to
the definition (2.3).

\qquad

\bigskip {\large 3. TEMPORAL FUNCTIONS}

The general solution of (2.8) or (2.11) can be presented as (superscripts
are omitted)

$S(\omega ,{\bf r})=S_{1}(\omega _{0},{\bf r})\exp \{i\int^{\omega }\tau
_{1}(\eta ,{\bf r})\ d\eta -\int^{\omega }\tau _{2}(\eta ,{\bf r})\ d\eta \}$%
, \qquad (3.1)

where low limits of integrals do not depend on $\omega $;

$\tau _{1}(\omega ,{\bf r})=\partial _{\omega }\arg S(\omega ,{\bf r})$
\qquad \qquad (3.2)

and

$\tau _{2}(\omega ,{\bf r})=\partial _{\omega }\ln |S(\omega ,{\bf r})|$
\qquad \qquad (3.3)

are, correspondingly, the Wigner-Smith formula of time delay at the process
of elastic scattering and the expression of extended duration of physical
state formation.

The unitarity of $S(\omega ,{\bf k})$ permits to conclude, with the
consideration of Cauchy- Schwartz inequality

$|S(\omega ,{\bf k})|^{2}\equiv 1=|\int S(\omega ,{\bf r})e^{-i{\bf kr}}d%
{\bf r}|^{2}\leq |\int S(\omega ,{\bf r})|^{2}d{\bf r}+\int \exp
\{-2\int^{\omega }\tau _{2}(\omega ,{\bf r})\ d\omega \}d{\bf r}$,(3.4)

that $\tau _{2}(\omega ,{\bf r})$ cannot retain the constant sign over all
frequencies interval. Its alternating may show an incompleteness of response
function in the given space point. It can be assumed that just $\tau _{2}$
must describe for the details of processes leading to the terminating of
reaction, to processes that are usually named as particles (states) dressing.

At the simplified consideration it can be concluded that as (2.16) at $\tau
_{2}\geq 0$ leads to the equality

$|R(\omega )|=|R(\omega _{0})|\exp [-(\omega -\omega _{0})\ \tau _{2}]$,
\qquad (3.4')

then the opportunity of Fourier-transformation of $R(\omega ,{\bf r})$, i.e.
the existence of the response function $R(t,{\bf r})$, dictate for the
considered theory the inequality:

$(\omega -\omega _{0})\tau _{2}(\omega _{0},{\bf r})\geq 0$. \qquad (3.5)

It shows that at $\omega <\omega _{0}$ the duration of formation $\tau
_{2}\leq 0$, i.e. in the certain frequencies range the advanced emission or
even superluminal phenomena are not excluded. Just such situation has place
at superluminal transfer of excitation and corresponds to a lot of
experimental data [22].

Addition of the following term of decomposition of $\ln R(\omega ,{\bf r}) $
to (2.4),

$\sigma (\omega ,{\bf r})\equiv -(\partial _{\omega })^{2}\ln R(\omega ,{\bf %
r})=-i\tau \prime (\omega ,{\bf r})$, \qquad (3.6)

at the inverse Fourier transformation with $\tau _{2}\geq 0$, i.e. for $%
\omega \geq \omega _{0}$, leads to the ''normal'' response function:

$R^{(+)}(t,{\bf r})=R(\omega _{0})(8\pi \sigma )^{-1/2}\exp \{-i\omega
_{0}t-(t-\tau )^{2}/2\sigma \}[1-$erf$((t-\tau )/\sqrt{2\sigma })]$. \qquad
(3.7)

This term shows the broadening of signals on their path.

With $\tau _{2}\leq 0$, i.e. at $\omega \leq \omega _{0}$, such
transformation results in the ''anomalous'' response function $R^{(-)}(t,%
{\bf r})$, which will be distinguished by the sign before the errors
function.

Thus, the complete response function is represented as the sum

$R(t)=\theta (\tau _{2})\ R^{(+)}(t)+\theta (-\tau _{2})\ R^{(-)}(t)$,
\qquad \qquad (3.8)

which can be examined as an analogue of decomposition of the causal
propagator $\Delta _{c}(x)=\theta (t)\Delta ^{(-)}+\theta (-t)\Delta ^{(+)}$%
, where $\Delta ^{(\pm )}$ propagators correspond to positive and negative
frequencies parts.

Let's consider some peculiarities of temporal functions connected with the
uncertainty principle.

It seems that the most general formal deduction of these relations was given
by Schr\"{o}dinger in [23]. The decomposition of the operators product on
the Hermitian and anti-Hermitian parts as ${\bf AB}=%
{\frac12}%
({\bf AB}+{\bf BA})+%
{\frac12}%
({\bf AB}-{\bf BA})$, subsequent quadrating of this expression, its
averaging over complete system of $\psi $-functions and replacement for
operators on difference of operators and their averaged values ${\bf A}%
\rightarrow \ {\bf A}-\left\langle {\bf A}\right\rangle $ bring to such
expression:

$(\Delta A)^{2}(\Delta B)^{2}\geq
{\frac12}%
|\left\langle {\bf AB}-{\bf BA}\right\rangle |^{2}+%
{\frac12}%
[\left\langle {\bf AB}+{\bf BA}\right\rangle -2\left\langle {\bf A}%
\right\rangle \left\langle {\bf B}\right\rangle ]^{2}$, \qquad \qquad (3.9)

which differs from the more usual form by the last term. The Heisenberg
limit of this expression shows a minimal value of uncertainties, which can
be achieved in the determined conditions.

The deduction of uncertainty principle with operator $\widehat{\tau }$ = $%
\partial $/i$\partial \omega $ was shown by Wigner in [24]. In considered
case the operators must be taken as $A\rightarrow E-\left\langle
E\right\rangle $ and $B\rightarrow t-\left\langle t\right\rangle $ and there
is needed the averaging (instead of $\psi $-functions) by the complete
system of $S(E)$ function, non-unitary in general, as $\left\langle {\bf A}%
\right\rangle =\int_{-\infty }^{\infty }dE\ S^{\ast }AS\ /\int dE\ |S|^{2}$.
The evident calculations give such result:

$(\Delta E)^{2}(\Delta t)^{2}\geq
{\frac14}%
\ \hbar ^{2}+%
{\frac14}%
\ [\left\langle E\tau _{1}\right\rangle -2\left\langle E\right\rangle
\left\langle \tau _{1}\right\rangle ]^{2}$, \qquad \qquad (3.9')

i.e. the general form of uncertainty does not depend on the formation
duration $\tau _{2}$ as it is the internal property of forming particle, but
can be enlarged by enlarging the duration of scattering process.

Note that this condition shows the possibility for enlarging the near field
extent. It seems that just this possibility is used in the near field
optics, where close to source are introducing additional macroscopic
scatterers and energy is varied by an external light flux [18].

As must be noted, the time domain processes are usually estimated via the
time-energy uncertainty principle $\Delta E\Delta t\geq
{\frac12}%
\hbar $. But in the cited article Wigner specially underlined that these
uncertainties depend on coordinates points, so if the process is progressing
in the $z$ direction:

$(\Delta t(z))^{2}=\int dxdydt\ (t-t_{0})^{2}|\psi (x,y,z,t)|^{2}/\int dydt\
|\psi (x,y,z,t)|^{2}$,

$(\Delta E(z))^{2}=\int dxdydE\ (E-E_{0})^{2}|\psi (x,y,z,E)|^{2}/\int dydE\
|\psi (x,y,z,E)|^{2}$ \qquad (3.10)

and can be different, in general case, for different $z$. This peculiarity
can be the starting point at investigation of phenomena of FTIR.

\qquad

\bigskip {\large 4. DISPERSION RELATIONS AND SUM RULES}

Response functions in the $(\omega ,{\bf r})$-representation obey the
temporal equation and simultaneously they are subject to the causality
principle, i.e. they are governed by the Kramers-Kr\"{o}nig dispersion
relations:

$S_{c}(\omega )=\frac{1}{\pi i}%
\displaystyle\int %
_{-\infty }^{\infty }\frac{d\eta }{\eta \ -\ \omega }S_{c}(\eta )$\qquad
\qquad (4.1)

(we write them in the simplest form with $S_{c}(\omega )\rightarrow 0$ at $%
\omega \rightarrow 0$). This duality permits to obtain the principal results.

By differentiation of (4.1) or by its substitution into (1.4) dispersion
relations can be represented in two forms:

$\tau (\omega )\ S_{c}(\omega )=-\frac{1}{\pi }%
\displaystyle\int %
_{-\infty }^{\infty }\frac{d\eta }{(\eta \ -\ \omega )^{2}}S_{c}(\eta )$,
\qquad \qquad (4.2)

$\tau (\omega )\ S_{c}(\omega )=\frac{1}{\pi i}%
\displaystyle\int %
_{-\infty }^{\infty }\frac{d\eta }{\eta \ -\ \omega }\ \tau (\eta )\
S_{c}(\eta )$. \qquad \qquad (4.2')

Equating of their right sides leads to the sum rule:

$%
\displaystyle\int %
_{-\infty }^{\infty }\frac{d\omega }{\omega }\ S_{c}(\omega )\ [\tau (\omega
)-\frac{i}{\omega }]=0$. \qquad \qquad (4.3)

This expression can be satisfied, in particular, with the equalities

$\tau _{1}(\omega )=0$, $\qquad \tau _{2}(\omega )=1/\omega $, \qquad \qquad
(4.4)

which show, and it is the principal conclusion, that even at the absence of
delay there is needed the certain time duration (twice bigger than the
uncertainties value) for formation of the out state (wave or particle, etc.).

Notice that from the temporal equation (1.4) at its Fourier transformation
follows also such expression:

$t\ S(t)=%
\displaystyle\int %
_{0}^{\infty }dt\prime \ S(t\prime )\ \tau (t-t\prime )$, \qquad $t\geq 0$.
\qquad \qquad (4.5)

Whether $\lim t\ S(t)=0$ at $t\rightarrow 0$, there must be rewarding the
sum rule:

$%
\displaystyle\int %
_{0}^{\infty }dt\ S(t)\ \tau (-t)$ $=0$. \qquad \qquad (4.5')

Further derivatives of the equation (1.4) lead to more complicate sum rules,
by checking of which can be determined the singularities of $S(t)$ at $%
t\rightarrow 0$.

As temporal functions $\tau (t)$ must be causal, there exist the independent
dispersion relations:

$\tau (\omega )=\frac{1}{\pi i}%
\displaystyle\int %
_{-\infty }^{\infty }\frac{d\eta }{\eta \ -\ \omega }\ \tau (\eta )$, \qquad
\qquad (4.6)

which evidently connect $\tau _{1}$ and $\tau _{2}$. They are consistent, in
particular, with the conditions (4.4) and with the representations of these
functions via propagators in the Section 6.

The analicity of causal response functions $S(\omega )$ permits to write
them in the form of Bl\"{a}schke product:

$S(\omega )=\ $const$\ \omega ^{-p}\prod_{n}\frac{\omega \ -\ \omega _{n}\
-\ i\gamma _{n}/2}{\omega \ -\ \omega _{n}\ +\ i\gamma _{n}/2}$. \qquad
\qquad (4.7)

With taking into account the relations (4.6) the sum rule (4.3) can be
rewritten via interaction operator $T(\omega )=i(S(\omega )-1)$ as:

$%
\displaystyle\int %
_{-\infty }^{\infty }d\omega \ \omega ^{-1}T(\omega )\ [\tau (\omega
)-i/\omega ]=0$. \qquad \qquad (4.3')

Since $\omega ^{p}S(\omega )$ is the meromorph function, the substituting of
$T(\omega )$ into this equality and closing the integration contour in the
upper half-plane produces the representation:

$\tau (\omega )=\sum_{n}1/[\omega -\omega _{n}+i\gamma _{n}/2]\pm \
ip/\omega $, \qquad $p>0$.\qquad \qquad\ (4.8)

Temporal functions have physical sense for positive frequencies, for
negative frequencies they are determined by the analytical continuation $%
\tau (-\omega )=\tau ^{\ast }(\omega )$, which follows from the analyticity
of $S(\omega )$. It permits the determination of Fourier transforms:

$\tau (t)=\frac{1}{2\pi i}%
\displaystyle\int %
_{-\infty }^{\infty }d\omega \ e^{i\omega t}\ \frac{\partial }{\partial
\omega }\ln S(\omega )=\sum $ res $e^{i\omega t}$, \qquad \qquad (4.9)

the last equality follows from the meromorphity of (4.7) at integer $p$. It
leads to the representations:

$\tau _{1}(t)=-\sum_{n}\cos (\omega _{n}t)\ \exp (-\gamma _{n}|t|)$; $\qquad
\tau _{2}(t)=-i\ $sgn$(t)$ $\tau _{1}(t)$,\qquad \qquad (4.10)

i.e. the temporal functions are represented by the set of damped oscillators
with self-frequencies modulated by the widths of transmission bands, where
index $n$ numerates self-frequencies.

It must be underlined the oscillator character (''time diffraction'') in the
expressions (4.10) and all oscillators are modulated by the self-band widths
(cf. [16]). Note, that at the standard approach the duration of processes
are usually taken as $1/\gamma $, without taking into account their
oscillation character.

The analyticity of $S(\omega +i\eta )$ in the upper half-plane permits to
write such integral over the closed contour:

$\oint \tau (\omega )\ d\omega =\oint \tau _{1}(\omega )\ d\omega =2\pi
(N-P) $, \qquad \qquad (4.11)

where $N$ and $P$ are zeros and poles of temporal function into the closed
contour. Poles of $\tau _{1}(\omega )$ signify impossibility of signal
transferring on these frequencies through the system (frequencies locking)
or particles capture at the scattering processes. Zeros show that
corresponding signals are passed via system without delays, etc. Really
(4.11) represents a variant of the Levinson theorem of quantum scattering
theory, e.g. [5].

The maximum-modulus principle for $|S(\omega )|$ shows, that as $\tau
_{2}(\omega )$ is determined via its derivative, it can not be equal to zero
at any frequency: the formation of outgoing signal (wave, particle, state)
always requires of some extended time duration.

It represents the main physical result of this Section.

\qquad

{\large 5. HARMONIC OSCILLATOR}

\qquad

Let's illustrate some of obtained results via consideration of the simplest
model, the oscillator with damping of (4.8)-type:

$\stackrel{\cdot \cdot }{x}-\ \gamma \stackrel{\cdot }{x}+\ \omega _{0}^{2}\
x=f(t)$. \qquad \qquad (5.1)

The complete causal solution of (5.1) can be written via the Green functions:

$x(t)=\int_{-\infty }^{t}dt\prime \ G(t-t\prime )\ f(t\prime )$;\qquad\ $%
G(t)=G_{0}(t)+G_{1}(t)$, \qquad \qquad (5.2)

and (5.2) can be considered as a model description of (2.13). The response
part of complete Green function is the solution of non-homogeneous equation,
Fourier image of which is

$G_{1}(\omega )=-1/2\pi (\omega -\omega _{1}+i\gamma /2)(\omega +\omega
_{1}+i\gamma /2)$ \qquad \qquad (5.3)

with $\omega _{1}^{2}=\omega _{0}^{2}-\gamma ^{2}/4$.

The corresponding causal temporal functions are:

$\tau _{1}(\omega )=\gamma /2[(\omega -\omega _{1})^{2}+\gamma
^{2}/4]+\{\omega _{1}\rightarrow -\omega _{1}\}$, \qquad \qquad (5.4)

$\tau _{2}(\omega )=(\omega -\omega _{1})/[(\omega -\omega _{1})^{2}+\gamma
^{2}/4]+\{\omega _{1}\rightarrow -\omega _{1}\}$. \qquad \qquad (5.5)

The last expression shows the possibility of advanced or superluminal
propagation at $\omega <\omega _{1}-\gamma ^{2}/8\omega _{1}$ (cf. [21] and
the superluminal transferring in macroscopic oscillator systems [25]).

Apart of some exotic cases $\gamma <<\omega _{0}$ and at $|\omega -\omega
_{0}|>\gamma $ and then at $\gamma \rightarrow 0$ it can be taken that

$\tau _{1}(\omega )\simeq \gamma /2[(\omega -\omega _{0})^{2}+\gamma
^{2}/4]\rightarrow \ \pi \delta (\omega -\omega _{0})$, \qquad \qquad (5.4')

$\tau _{2}(\omega )\simeq (\omega -\omega _{1})/[(\omega -\omega
_{1})^{2}+\gamma ^{2}/4]\rightarrow \ 1/(\omega -\omega _{1})$, \qquad
\qquad (5.5')

which shows the proximity of last expression to the uncertainty values, but
it must be specially underlined its twice bigger numerical value. It means
possibility of measurements of these values and therefore the observability
of connected phenomena.

It seems that the most evident and close to the intuitive physical
representation of temporal functions may give their description in the
Lorentz model of dispersing and absorbing media (e.g. [26]), where media are
described as the set of oscillators with damping. Each oscillator is
describing by the Green function (5.3) with corresponding factor depending
on density of scatterers, etc.

The real part of dielectric susceptibility and conductance are expressed in
this model, respectively, as

$\varepsilon _{1}(\omega )-1\simeq \omega _{p}^{2}\ (\omega _{0}-\omega
)/2\omega \lbrack (\omega _{0}-\omega )^{2}+\gamma ^{2}/4]$; \qquad \qquad
(5.6)

$\sigma _{el}(\omega )\simeq \omega _{p}^{2}/8\pi \gamma \ [(\omega
_{0}-\omega )^{2}+\gamma ^{2}/4]$, \qquad \qquad (5.7)

$\omega _{p}\ $is the plasma frequency.

The comparison of (5.6-7) to (5.4-5) suggest, excluding the nearest vicinity
of resonance, the possibilities of approximations:

$\varepsilon _{1}(\omega )-1\simeq (\omega _{p}^{2}/2\omega )\ \tau
_{2}(\omega )$, \qquad \qquad (5.6')

$\sigma _{el}(\omega )\simeq (\omega _{p}^{2}/4\pi \gamma ^{2})$\ $\tau
_{1}(\omega )$. \qquad \qquad (5.7')

These relations give the evident interpretation of both temporal functions.
So the polarization of media is reasonably determined by durations of waves
formation. And, as it is also intuitively evident, the electrical
conductivity, as (every) transfer process, is determined via the durations
of EM waves delay, which can be induced by virtual moment transfers between
charged particles, i.e. by their retarded movements in the EM flux direction.

The more general connection of temporal functions with characteristics of
media can be established in such fashion. The principle of entropy grows
requires of execution of the strong inequality for almost transparent
passive dispersing media: $\partial (\omega \varepsilon )/\partial \omega
\geq 0$ [27]. With the substitution $R\rightarrow \varepsilon (\omega
)-\varepsilon (\infty )=\varepsilon _{1}+i\varepsilon _{2}$, i.e. by the
equation $\partial \varepsilon /\partial \omega =i\ \tau \ \varepsilon $,
the real part of this general inequality is rewritten as

$\tau _{2}\leq 1/\omega -\tau _{1}\varepsilon _{2}/\varepsilon _{1}$, \qquad
\qquad (5.8)

As for sufficiently low frequencies $\varepsilon _{2}=(4\pi /\omega )\
\sigma _{el}(\omega )$, this inequality reduces to the simplest form:

$\tau _{1}+\tau _{2}\leq 1/\omega $, \qquad \qquad (5.9)

which evidently show that $\tau _{2}$ can be negative in some frequencies
regions. In particular it must be negative in the region of anomalous
dispersion, where must be expected a discordance between maxima of $\tau
_{1} $ and $\tau _{2}$ [21], but for their description are needed more
complicate models.\qquad

\qquad

{\large 6. TEMPORAL WIGNER FUNCTIONS}

\qquad

As the cited Frank theory explains the \^{C}erenkov photons emission via
interference of classical waves emitted at different points, this effect in
the quantum theory would be interpreted via the Wigner functions that just
describe the overlapping of space domains of states [28]:

$w({\bf k};{\bf r};t)=(\frac{1}{2\pi })^{3}%
\displaystyle\int %
d{\bf q\ }e^{i{\bf qk}}\psi ({\bf r}-{\bf q}/2;t)\ \psi ^{\ast }({\bf r}+%
{\bf q}/2;t)$, \qquad \qquad (6.1)

or via their covariant generalization [29]:

$w(k;x)=(\frac{1}{2\pi })^{4}%
\displaystyle\int %
dv\ e^{-ivk}\psi (x-v/2)\ \psi ^{\ast }(x+v/2)$, \qquad \qquad (6.2)

with 4-vectors $k$, $x$, $v$ that describe the time-space overlapping
(interference) of the quantum self-states. The quantum field interpretation
of (6.2) through the creation and destruction operators descriptively shows
that the interference of oppositely shifted wave functions in it must sum
the maps of their possible variation onto 4-intervals.

Let us consider the one-particle temporal Wigner functions as the special
case of (6.2),

$w^{(+)}(\omega ,t;{\bf r})=\frac{1}{2\pi }%
\displaystyle\int %
_{0}^{\infty }d\tau \ e^{i\omega \tau }\psi (t-\tau /2;{\bf r})\ \psi ^{\ast
}(t+\tau /2;{\bf r})$; \qquad \qquad (6.3)

$w^{(-)}(\omega ,t;{\bf r})=w^{(+)}(-\omega ,t;{\bf r})$. \qquad \qquad
(6.3')

These functions evidently describe the overlap of time-shifted wave
functions at one space point and therefore just these functions should
characterize the time delay at collision process and the duration of states
formation (space arguments will be hereafter omitted).

By time shifts of wave functions with the Hamiltonian ${\bf H}$,

$\psi (t-\tau /2)=\psi (t)\ \exp (i{\bf H}\tau /2)$;

$\psi ^{\ast }(t+t/2)=\exp (i{\bf H}\tau /2)\ \psi ^{\ast }(t)$, \qquad
\qquad (6.4)

the temporal Wigner function (6.3) is rewritten as

$w^{(+)}(\omega ,t)=\psi (t)\ \delta _{+}(\omega -{\bf H})\ \psi ^{\ast
}(t)\rightarrow \ \psi (t)\ W^{(+)}(\omega ,t)\ \psi ^{\ast }(t)$. \qquad
\qquad (6.5)

These functions are the self-functions of the operator equation

$\frac{\partial }{i\partial \omega }\ w^{(+)}(\omega ,t)=i(\omega
-E)^{-1}w^{(+)}(\omega ,t)$ \qquad \qquad (6.6)

of the (1.4) type, where $E$ is the (complex) energy of system, ${\bf H}\psi
=E\psi $. This equation can be considered as the reciprocal one to the
Liouville equation in the Schr\"{o}dinger representation. It shows that the
durations of scattering processes and of states formation should be
described as the self-values of corresponding Green operators.

It must be noted that in distinction from the space Wigner functions the
temporal functions are non-symmetric relative to their parameters and
therefore their self-values can be complex ones. It just corresponds to
possibilities of retarded and advanced interactions.

Slightly another derivation of such equation can be examined on transition
to the Heisenberg representation,

$\psi (t-\tau /2)=\exp (i\omega \widehat{{\bf \tau }})\ \psi (t/2)\ \exp
(-i\omega \widehat{{\bf \tau }})$,

$\psi ^{+}(t+\tau /2)=\exp (i\omega \widehat{{\bf \tau }})\ \psi ^{+}(t/2)\
\exp (-i\omega \widehat{{\bf \tau }})$, \qquad \qquad (6.4')

with a temporal operator $\widehat{{\bf \tau }}$. The equation (6.6) can be
rewritten as

$-i\partial _{\omega }w^{(+)}(\omega ,t)=\exp (i\omega \widehat{{\bf \tau }}%
)\ [\widehat{{\bf \tau }},Q]\ \exp (-i\omega \widehat{{\bf \tau }})$, \qquad
\qquad (6.6')

with function

$Q(\omega )=\frac{1}{2\pi }%
\displaystyle\int %
_{0}^{\infty }dt\ e^{i\omega \tau }\psi (-\tau /2)\ \psi ^{+}(\tau /2)$.

This representation naturally leads to the Hamilton equations for temporal
operators.

The function (6.3), just as the Wigner functions, can be rewritten via the
conjugate variable, via the energy shifts,

$-i\partial _{\omega }w^{(+)}(\omega ,t;{\bf r})=\frac{1}{2\pi }%
\displaystyle\int %
_{0}^{\infty }d\eta \ e^{i\eta t}\ \psi (\omega -\eta /2;{\bf r})\ \psi
^{+}(\omega +\eta /2;{\bf r})$.\qquad\ \qquad (6.7)

Therefore the state formation can be considered as a gradually process of
variation of energy till their definite values for physical (''dressed'')
particles. This property can be evidently generalized on interactions of
arbitrary number of particles. In an analogical way may be considered the
gradual evolution of (establishment in) other particles characteristics in
the processes of interaction

It can be noted, in particular, that if it is possible to introduce the
operator of complete moment ${\bf K}$, the Wigner functions in the close
analogy with all above can be symbolically written as

$w({\bf k};{\bf r})=\psi ({\bf r})\ \delta ({\bf k}-{\bf K})\ \psi ^{+}({\bf %
r})$; \qquad \qquad (6.8)

i.e. via the vector Green functions. (This possibility will not be
considered here further.)

\qquad

\bigskip {\large 7. FORMAL THEORY OF SCATTERING}

\qquad

Let's consider more scrupulously the representation of temporal functions
via propagators for the process of elastic scattering:

$a+b\rightarrow \ a+b$.\qquad \qquad \qquad\ (7.1)

The kinetics of interaction must be described by the operator ${\bf S}=1-i%
{\bf T}$, where ${\bf T}$ is the operator of interaction, expressed via
propagators $G(E)=(E-{\bf H})^{-1}$ and $g(E)=(E-{\bf H}_{0})^{-1}$, the
complete Hamiltonian ${\bf H}={\bf H}_{0}-{\bf V}$, self-values of the
Hamiltonians are complex, ${\bf H}\psi $ = $(E+i\Gamma )\psi $ and ${\bf H}%
_{0}\psi _{0}$ = $(E_{0}+i\Gamma _{0})\psi _{0}$, where $\Gamma _{0}$ and $%
\Gamma $ are the natural and complete widths of the upper level (for the
sake of simplicity there is considered the simplest two-level system).

As it was shown in [7] the duration of scattering and duration of newly
state formation are naturally expressed via propagators with account and
without account of this interaction:

$\Delta \widehat{\tau }\equiv \widehat{\tau }-\widehat{\tau }%
_{0}=i[G(E)-g(E)]$, \qquad \qquad (7.2)

where $\widehat{\tau }$ and $\widehat{\tau }_{0}$ denote temporal
characteristics of complete particle path with and without interaction.

The expression (7.2) follows from differentiation of the operator of
interaction, ${\bf T}={\bf V}/(1-g{\bf V})$, with taking into account the
Dyson equation $G=g+g{\bf V}G$ and the definition of temporal operator (7.2)
via equation

$\partial {\bf T}/i\partial E=\Delta \widehat{\tau }\ {\bf T}$. \qquad
\qquad (7.3)

Under the transition to energy surface, $E=E({\bf p})$, the matrix element
of (7.2),

$\left\langle {\bf p}|\Delta \widehat{\tau }|{\bf p}\right\rangle =\pm \ i%
\mathop{\displaystyle\sum}%
\{[E-E_{n}-i\Gamma _{n}/2]^{-1}-[E-E_{n(o)}-i\Gamma _{n(o)}/2]^{-1}\}$,
\qquad \qquad (7.4)

clearly shows its properties. So $iG(E)$ can be interpreted as the time
duration needed for particles flight and their elastic scattering and $ig(E)$
corresponds to the free transfer.

Transition in (7.2) into the coordinate representation,

$G({\bf r})-g({\bf r})=-\frac{1}{(2\pi )^{3}}%
\displaystyle\int %
d{\bf p\ }\left\langle {\bf p}|\Delta \widehat{\tau }|{\bf p}\right\rangle \
e^{i{\bf pr}}$, \qquad \qquad (7.5)

demonstrates the similarity of our definition with the Smith derivation of
time delay at scattering processes [4].

Notice that from (7.2) follows such expression for the temporal operator:

$\Delta \widehat{\tau }$ $=ig\ {\bf V\ }G$, \qquad \qquad (7.6)

which permits, in particular, the expansion of temporal functions into the
series about the free Green functions and interaction vertices:

$\Delta \widehat{\tau }=ig{\bf V}g+ig{\bf V}g{\bf V}g+...$ , \qquad \qquad
(7.7)

natural for quantum theories and useful for interpretations of these
processes via Feynman graphs, etc. These forms show that the measurement of
time characteristics of process is equivalent to addition of specific vertex
(vertices) to corresponding diagrams of process (we shall return to this
interpretation below).

The third form of temporal operator, which follows from (7.2), can be
expressed as

$\Delta \widehat{\tau }=ig{\bf T}g$. \qquad \qquad (7.6)

Its matrix element,

$i\left\langle {\bf p}|{\bf T}|{\bf p}\right\rangle /[(E-E_{0}({\bf p}%
))^{2}+\Gamma _{0}^{2}({\bf p})/4]$, \qquad \qquad (7.7)

by the substitution of the known expression for scattering amplitude on the
angle zero, $f({\bf p},{\bf p})=4\pi ^{2}m\ \left\langle {\bf p}|{\bf T}(%
{\bf p})|{\bf p}\right\rangle $, and the transferring to energy surface $E=E(%
{\bf p})$ leads to the expression:

$\left\langle {\bf p}|\Delta \widehat{\tau }|{\bf p}\right\rangle =\frac{1}{%
2\pi ^{2}im\Gamma ^{2}}$ $f({\bf p},{\bf p})$.\qquad\ \qquad (7.8)

The real part of (7.8) can be expressed, with taking into account the
optical theorem of scattering theory, via the total cross-section of
scattering:

$\tau _{1}({\bf p})=\frac{p}{(2\pi )^{3}m\Gamma ^{2}}\ \sigma _{tot}({\bf p}%
) $. \qquad \qquad (7.9)

Just this result clarifies the great delay with the beginning of
investigation of temporal characteristics of scattering processes: the most
part of this information is contained in the Green functions and
cross-sections.

If we determine the volume of interaction as $V=\sigma _{max}u\ \tau _{max}$%
, where $u$ is the velocity of scattered particle, $\tau _{max}=2/\Gamma $
and $\sigma _{max}$ is the resonance cross-section, the mean value of
duration of interaction can be determined as the balance relation,

$\overline{\tau _{1}}({\bf p})=\sigma _{tot}({\bf p})\ \tau _{max}/\sigma
_{max}$. \qquad \qquad (7.10)

For the most practically important optical region $\Gamma \ \symbol{126}\
10^{8}\sec ^{-1}$, $\sigma _{max}=4\pi /k^{2}$, $\sigma _{tot}=(4\pi /k)\
r_{0}$, $r_{0}=e^{2}/mc^{2}$. Therefore for $k$ $=6.3\cdot (10^{4}\div
10^{3})$ cm$^{-1}$ the expression (7.10) leads for nonresonant frequencies to

$\overline{\tau _{1}}({\bf p})\ \symbol{126}\ (k/\Gamma )\ r_{0}=1.6\cdot
(10^{-16}\div 10^{-15})\ \sec $, \qquad \qquad (7.11)

which evidently corresponds to the observable data.

It can be shown that this value permits to estimate the mean value of index
of refraction in nonresonant region. As it had been shown in [30] the
optical dispersion in the transparent, at least, region can be considered as
the kinetic process of photons transfer through media. Such transfer must be
described for the free path lengths $\ell =1/N\sigma _{tot}$ with the vacuum
velocity $c$, where$\ N$ is the density of outer (optical) electrons, and
the subsequent delays at each scattering for the mean time of order (7.11).
So, the complete time, needed for photons transfer on distance $L$, is equal
to

$\overline{T}=(L/c)+(L/\ell )\ \tau _{1}$. \qquad \qquad (7.12)

This estimation leads to the group velocity $u=L/\overline{T}$ and, for
nonresonant cases, to the group index of refraction:

$n_{gr}\equiv \frac{c}{u}=\frac{c\overline{T}}{L}=1+cN\sigma _{tot}\tau
_{1}\ \symbol{126}\ 1+N\frac{4\pi c}{\Gamma }r_{0}^{2}\ \symbol{126}\
1+3\times 10^{-22}N$, \qquad \qquad (7.13)

which qualitatively corresponds to the observations ($N$ is of order of the
L\"{o}schmidt number).

It must be underlined that the representation of temporal functions via
propagators supports the results of the Section 6: their analytical
properties and the existence of dispersion relations of Kramers-Kr\"{o}nig
type (cf. the estimation of such relations with possible subtractions,
connected with renormalization procedures [31]).

\qquad

{\large 8. DURATION OF INTERACTION AND ADIABATIC HYPOTHESIS}

\qquad

Let us show that the magnitudes of duration of interaction are implicitly
contained in the standard theory in the form of adiabatic hypothesis. This
hypothesis asserts that for the correct quantum calculations of transition
amplitude there is needed such artificial substitution for the Hamiltonian:

$V(t)\rightarrow \ V(t)\exp (-\lambda |t|)$ \qquad \qquad (8.1)

with passage to the limit $\lambda \rightarrow 0$ after all calculations
(e.g. [5]).

Stueckelberg proposed more general approach to these problems via the
causality condition [32]. Bogoliubov generalized this method by introduction
of operations of ''the switching interaction on and off'', i.e. of some
function $q(x)\in \lbrack 0,1]$, which characterizes the intensity of
interaction: in the space-time regions with $q(x)=0$ interaction is
completely absent and at $q(x)=1$ is completely switched [33]. But the
introduction of this switching function has not physical substantiation and
can be justified a posteriori only.

In this theory $S$-matrix becomes a functional of function $q(x)$ and the
final state of system in the interaction representation is expressed as

$\Phi \lbrack q]=S[q]\ \Phi _{0}$, \qquad \qquad (8.2)

where $\Phi _{0}$ is the initial state. For performing of this program the
switching function is introduced into the (classical) action function, e.g.

$S_{cl}=\int dx\ q(x)\ \check{L}(x)$, \qquad \qquad (8.3)

where $\check{L}(x)$ is the density of Lagrangian of interaction. In the
quantum field theory, correspondingly, the operator of evolution will be
represented as the functional:

$S[q]=T^{\prime }\ \exp \{i\int dx\ q(x)\ \check{L}(x,q)\}$, \qquad \qquad
(8.4)

$T^{\prime }$\ is the chronologization operator and it is assumed that the
relative value of Lagrangian depends on ''intensity of interaction''.

The variation of (8.2) over $q(x)$ leads to the variational equation

$i\delta \Phi \lbrack q]/\delta q(x)={\bf H}(x;q)\ \Phi (q)$ \qquad \qquad
(8.5)

with the Hamiltonian of interaction

${\bf H}(x;q)=i(\delta S[q]/\delta q(x))\ S^{\ast }[q]$, \qquad \qquad (8.6)

which is the evident variational analog, at $q=1$, of the Schr\"{o}dinger
equation for $S$-operator in the interaction representation. This form leads
to the covariant Tomonaga-Schwinger equation.

The switching function $q(x)$ describes the 4-region of interaction, and if
we shall {\it assume} that the magnitude of this region depends on details
of interaction, we rewrite (8.4) as

$S[\check{L}]=T^{\prime }\exp \{-i\int dx\ q(x,\check{L})\ \check{L}%
(x)\}=T^{\prime }\exp \{-i\int dk\ q(-k,\check{L})\ \check{L}(k)\}$, \qquad
\qquad (8.7)

in the last equality the existence of corresponding Fourier transforms is
proposed. This transition from (8.4) to (8.7) can be considered as the
Legendre-type transformation $q\longleftrightarrow \check{L}$ of the
classical action function (8.3), i.e. instead of consideration of switching
of intensity of interaction there is considered a variable part of 4-volume
of interaction (in particular, of the duration of interaction). This
assumption can be also justified only a posteriori.

Thus we vary (8.7) over $\check{L}(k)$ and it leads to the equation

$\delta S[\check{L}]/i\delta \check{L}(k)=q(-k,\check{L})\ S[\check{L}]$,
\qquad \qquad (8.8)

or

$q(-k,\check{L})=(\delta S[\check{L}]/i\delta \check{L}(k))\ S^{-1}[\check{L}%
]$, \qquad \qquad (8.8')

i.e. to the evident variation-type analog of the equation for temporal
operator. Notice that in the complete accordance with the Bogoliubov method
it can be considered the singularity of $\check{L}$ on a hypersurface $%
\sigma (\omega )$, which would lead to the equation

$\delta S/i\delta \check{L}(k,\sigma )=q(-k,\sigma )\ S(\sigma )$, \qquad
\qquad (8.9)

reciprocal to the Tomonaga-Schwinger equation.

From these equations naturally follows the equation (1.4), reciprocal to the
Schr\"{o}dinger equation for $S$-matrix, with the formal temporal function

$\tau (\omega )=\int dk\ q(-k,\check{L})\ (\delta \check{L}(k)/\delta \omega
)$. \qquad \qquad (8.9)

The switching function $q(x)$ can be presented, in accordance with the
adiabatic hypothesis (8.1), as

$q(x)=\exp (-\gamma |t|/2)$ \qquad or $\qquad q(-k)=\delta (k)/2\pi
i(k_{0}\pm \ i\gamma /2)$. \qquad \qquad (8.10)

These expressions can be rewritten in the covariant form by introduction of
any unit time-like vector $n_{\mu }$ and replacement for $\tau $ $%
\rightarrow $\ $n_{\mu }x_{\mu }$. The substitution of (8.10) into the
expression (8.9) with assuming of the $\delta $-type properties of $\delta
\check{L}(k)/\delta \omega $ and with the frequency's shift $\omega
_{0}\rightarrow \omega -\omega _{0}$, leads to the usual form of temporal
function for the simplest two-level system,

$\tau \equiv \tau _{1}+i\tau _{2}=1/\pi (\gamma /2\pm \ i(\omega -\omega
_{0}))$. \qquad \qquad (8.11)

Thus it can be concluded that the adiabatic hypothesis presents a
non-obvious introduction of the time duration concept in the theory.

\qquad

{\large 9. QUANTUM ELECTRODYNAMICS}

\qquad

Let us begin the consideration of temporal functions of QED with examination
of the photon causal propagator of lowest order in vacuum (Feynman
calibration, $\eta \rightarrow 0+$):

$D_{c}(\omega ,{\bf k})=4\pi /(\omega ^{2}-{\bf k}^{2}+i\eta )$. \qquad
\qquad (9.1)

In accordance with all above it conducts to such expressions for time delay
and duration of formation:

$\tau _{1}=-2\pi \ \delta (\omega ^{2}-{\bf k}^{2})$, \qquad \qquad (9.2)

$\tau _{2}=2\omega /(\omega ^{2}-{\bf k}^{2})\ \symbol{126}\ 1/(\omega -|%
{\bf k}|)$. \qquad \qquad (9.3)

The function $\tau _{1}$ simply shows that the photon can be absorbed or
emitted only completely. The function $\tau _{2}$ qualitatively corresponds
to the uncertainty principle, but is twice bigger, i.e. this time can be
measurable; it shows the possibility of retarded, at $\omega >|{\bf k}|$, or
advanced, at $\omega <|{\bf k}|$, emission of photon.

It should be noted that the consideration of complete propagators through
replacements ${\bf k}^{2}\rightarrow {\bf k}^{2}+P(k)$ in (9.1) with the
polarization operator $P(k)$ of QED or even transition to propagators of
massive (scalar, for simplicity) particles does not change these general
results.

The estimations of temporal values for elementary processes in the nearest
orders can be achieved by such simple procedure: in accordance with (3.3) it
can be suggested the expression for $\tau _{2}$ via cross-section of
scattering:

$\tau _{2}=-%
{\frac12}%
(\partial /\partial \omega )\ln \sigma $. \qquad \qquad (9.4)

So as for the Rutherford scattering $\sigma \ \symbol{126}\ E^{-2}$, it
gives, in accordance with (9.3), $\tau _{2}=1/E$; for the nonrelativistic
limit of Compton scattering $\sigma \ \symbol{126}\ (1-2\omega /m)$ and
therefore $\tau _{2}=1/m\ (1-2\omega /m)$, etc. The values of $\tau _{1}$
can be estimated now via dispersion relations (4.6) and so on.

The complete covariant generalization of the temporal operator $\widehat{%
\tau }$ can be achieved by the Legendre transformation of equations for the
4-moments of interaction:

$i\partial S/\partial x_{\mu }=k_{\mu }\ S$ \qquad $\longleftrightarrow
\qquad $ $\partial S/i\partial k_{\mu }=x_{\mu }\ S$, \qquad \qquad (9.5)

where $x_{\mu }=(t,{\bf r})$ represents the 4-vector of ''duration-space
extent of interaction'' in (2.12). Notice that for investigation of some
process the substitution $S\rightarrow M$, i.e. the consideration of
concrete matrix element of the process is implied.

The time operator is now generalized as the covariant operator $\partial
/i\partial p_{\mu }$, canonically conjugated to the energy-momentum operator
$i\partial /\partial x_{\mu }$. The determination of corresponding functions
can be established via the Ward-Takahashi identity:

$\partial G/i\partial p_{\mu }=G(p)\ \Gamma _{\mu }(p,p;0)\ G(p)$, \qquad
\qquad (9.6)

$G(p)$ is the particle Green function, $\Gamma _{\mu }(p,q;p-q)$ is the
vertex part. From it follows the expression for self-values of 4-operator:

$\xi _{\mu }(p)\equiv \partial \ln G/i\partial p_{\mu }=%
{\frac12}%
\{G(p)\ \Gamma _{\mu }(p,p;0)+\Gamma _{\mu }(p,p;0)\ G(p)\}$.\qquad\ \qquad
(9.6')

The 4-vector $\xi _{\mu }$ consists of the temporal and space components, $%
\xi _{0}(p)\equiv \tau (\omega ,{\bf k})$ and ${\bf \xi }(p)\equiv {\bf \rho
}(\omega ,{\bf k})$, which determines the space extents of interaction.
(Similar operators were introduced for localized states of spin zero massive
particles [34], but they are a matter of discussions for photons [35].)

The known representation of vertex operator $\Gamma _{\mu }(p,p;0)=\gamma
_{\mu }-(\partial /\partial p_{\mu })\Sigma $ with the mass operator $\Sigma
$ shows that the expression (9.6) is connected with the extended formation
of physical particles parameters.

The difference between both parts of $\xi _{\mu }$ can be demonstrated by
consideration of the simplest case, the complete causal propagator $D_{c}=%
\overline{D}+D_{1}$ in the scope of scalar electrodynamics. In accordance
with (9.6') both parts of temporal function in the $p$-representation are
equal to

$\xi _{\mu 1}(p)\equiv \ $Re$\ \xi _{\mu }(p)=p_{\mu }\ D_{1}(p;m)=p_{\mu
}(D^{(+)}-D^{(-)})$, \qquad \qquad (9.7)

$\xi _{\mu 2}(p)\equiv \ $Im$\ \xi _{\mu }(p)=p_{\mu }\ \overline{D}%
(p;m)=p_{\mu }(D_{ret}-D_{adv})$. \qquad \qquad (9.7')

In the $x$-representation these expressions are even more descriptive:

$\xi _{\mu 1}(x)=(\partial /\partial x_{\mu })(D^{(+)}(x)-D^{(-)}(x))$;
\qquad \qquad (9.8)

$\xi _{\mu 2}(x)=(\partial /\partial x_{\mu })(D_{ret}(x)-D_{adv}(x))$,
\qquad \qquad (9.8')

i.e. the duration of interaction describes the decreasing of
negative-frequency part and increasing of positive-frequency part, the
extended duration of state formation determine the difference of retarded
and advanced parts alteration.

These expressions evidently show also the difference between uncertainty
values and durations or space extents of interactions. So from the
expression (9.7') and as $\overline{D}(p)=-$P$\frac{1}{k^{2}}$ follows that $%
\tau _{2}$ and ${\bf \rho }_{2}$ are approximately twice bigger than the
corresponding uncertainty values.

Notice that these representations can lead to several particular models.
Let's consider as example the space extent of particle formation averaged
over frequencies not excided its rest mass:

$\left\langle \overline{\Delta }({\bf r},m)\right\rangle =\frac{1}{m}%
\displaystyle\int %
_{0}^{m}d\mu \ \overline{\Delta }({\bf r},\mu )=\sin (mr)/4\pi mr^{2}$,

its gradient describes, via (9.8'), the space extent of interaction, and it
approaches, in accordance with the uncertainty principle, to $\delta ({\bf r}%
)$ with increasing of the mass of particle.

The temporal functions for electron must be determined via the electron
Green functions and in the nearest order they are represented through (9.6')
as

$\left\langle \tau (p)\right\rangle =%
{\frac12}%
$Tr$\{\gamma _{0}\ S_{(.)}(p)\}=p_{0}\Delta _{(.)}(p)$, \qquad \qquad (9.9)

which at the substitutions for $\omega \rightarrow (p_{0}^{2}-{\bf p}%
^{2})^{1/2}$ and the Fourier transformation over moments variables coincides
with (9.2-3).

The physical sense of these functions can be established in such a way. The
expression (9.8) shows that the temporal measurement, for which $\mu =0$, is
equivalent to addition of zero-frequency scalar photon line to the
appropriate electron lines of the Feynman graphs. Therefore the durations
can be interpreted via the probed additional Coulomb fields of zero
intensity (cf. with the Baz' method of zero-intensity probe magnetic field
and the ''Larmor clock'' in it [9, 10]).

This examination demonstrates, in particular, that the superluminal
phenomena may be observed, in principle, into all scattering processes, and
not only in processes of QED.

For the spinor quantum electrodynamics this 4-vector must be determined,
correspondingly, as

$\xi _{\mu }=\ $Tr$\ (M^{+}\frac{\partial }{i\partial k_{\mu }}M)\ /\ $Tr$\
(M^{+}M)$. \qquad (9.10)

It seems interesting to check by this expression the results, obtained in
[11 - 14] for bremsstralung. By insertion of its known matrix element (i.e.
[36]) into (9.10) it can be easy shown that Re$\ \xi _{\mu }=0$ in the
lowest order. It corresponds to the absence of any delay at bremsstralung,
but the components of Im $\xi _{\mu }$ , which are connected to the
formation processes, are nonzero. So if $\epsilon $, ${\bf k}$ and $\epsilon
^{\prime }$, ${\bf k}^{\prime }$\ are the initial and final electron
energies and moments, $\omega $ is the photon energy, $\vartheta $ is the
angle of electron departure, we have that the duration and corresponding
path extent of electron dressing are determined as

$\tau _{2}=\frac{1}{\omega }$, \qquad $\rho _{2}=\frac{k^{\prime }}{\epsilon
\omega }+%
{\frac12}%
\frac{k^{\prime }\ -\ k}{\epsilon \epsilon ^{\prime }\ +\ m^{2}}$ \qquad
\qquad (9.11)

at $\epsilon ,\epsilon ^{\prime }\ \geq m$ and

$\tau _{2}\approx |\rho _{2}|\ \symbol{126}\ \frac{2\epsilon (\epsilon
^{\prime }\ +\ \omega )}{m^{2}\omega }$, \qquad $\rho _{\bot }\symbol{126}\
\frac{2\epsilon \vartheta }{m^{2}}$, \qquad \qquad (9.12)

when $\epsilon ,\epsilon ^{\prime }\gg m$.

These results correspond to the previous calculations, but are obtained by a
shorter and more general way. Notice that the region of photons formation
can be considered as the near field of classical electrodynamics.

Let's consider shortly, as an example, some more general problems.

So if we investigate the scattering of scalar particles via the one-particle
exchange, the lowest order values of $\xi _{\mu }$ are determined as the
logarithmic derivatives of intermediate particle propagator. In the standard
notation with taking into account the Ward-Takahashi identity it leads to
the expression,

$\xi _{\mu }(k)=(\partial /i\partial k_{\mu })\ln D_{c}^{\prime }\ =[2k_{\mu
}/i(t-m^{2})^{2}]\ \Gamma (t,t,0)\ D_{c}^{\prime }$, \qquad \qquad (9.13)

where $D_{c}^{\prime }$\ is the complete Green function.

The factor $2k_{\mu }/(t-m^{2})$, that formally is close to the uncertainty
principle, corresponds to the duration of outgoing particles formation $\tau
_{2}=|\rho _{2}|\ \symbol{126}\ 1/2E$. Time delay is connected with the
imaginary part of propagator and arises at $t\geq 4m^{2}$, i.e. with
possibilities of new particles birth.

As well as under photons formation the length of their formation (the near
field region) appears, it can be proposed that in processes of formation of
particles with additional internal parameters would manifest itself another
regions of their formation with their own peculiarities.

\qquad

{\large 10. TO INTERPRETATION OF SOME RENORMALIZATION PROCEDURES}

\qquad

Let's begin with the Pauli-Villars method of regularization.

This method consists in such substitutions:

$\Delta (p,m)\rightarrow \ \Delta (p,m)-\Delta (p,M)\ \symbol{126}\ \frac{%
(m^{2}\ -\ M^{2})}{(p^{2}\ -\ m^{2}\ +\ i\eta )(p^{2}\ -\ M^{2}\ +\ i\eta )}$
\qquad \qquad (10.1)

with further passage to the limit $M\rightarrow \infty $.

What is its physical sense? Such substitutions implies a decreasing of
duration of new state formation with $p^{2}<M^{2}$:

$\xi _{\mu 2}\ \symbol{126}\ 2p_{\mu
}\{(p^{2}-m^{2})^{-1}+(p^{2}-M^{2})^{-1}\}$, \qquad \qquad (10.2)

i.e. it seems as a procedure of alteration of the interaction 4-volume,
which was discussed in connection with the adiabatic hypothesis.

In the $x$-representation such substitution, $\xi _{\mu }(x)\rightarrow \
(\partial /\partial x_{\mu })\{\Delta _{c}(x,m)+\Delta _{c}(x,M)\}$, leads
to increasing of the role of more energetic and more deep-seated virtual
excitations at the beginning of calculations. And it actually means a
partial account of higher terms of $S$-matrix in the process of particle
formation.

Let's pass on to the subtraction procedures of renormalization.

The regularized mass function of the electron propagator is determined as

$\Sigma ^{reg}(p)=\Sigma (p)-\Sigma (p)\left| _{\gamma p=m}\right. -(\gamma
p-m)\ (\partial \Sigma (p)/\partial p)\left| _{\gamma p=m}\right. $. \qquad
\qquad (10.3)

Then the equalities

$\Sigma ^{reg}(p)\left| _{\gamma p=m}\right. =0$, $\qquad (\partial \Sigma
^{reg}(p)/\partial p)\left| _{\gamma p=m}\right. =0$, \qquad \qquad (10.4)

postulated for its renormalization, now can be interpreted as the physically
justified conditions: the mass of particle has definite magnitude and the
process of its accumulation should be finished at the finite time.

The regularized self-energetic part of the photon propagator ($k\partial
_{k}\equiv k_{\nu }\partial /\partial k_{\nu }$)

$\Pi _{\mu \nu }^{reg}(k)=\Pi _{\mu \nu }(k)-\{1-k\partial _{k}-%
{\frac12}%
(k\partial _{k})^{2}\}\Pi _{\mu \nu }(k)\left| _{k=0}\right. $. \qquad
\qquad (10.5)

Apart from the evident and gauge equalities,

$\Pi _{\mu \nu }^{reg}(0)=0$; \qquad $k^{-2}\Pi _{\mu \nu }^{reg}(k)\left|
_{k=0}\right. =0$, \qquad \qquad (10.6)

the conditions, that are usually simply postulated:

$k\partial _{k}\Pi _{\mu \nu }^{reg}(k)\left| _{k=0}\right. =0$, $\qquad
(k\partial _{k})^{2}\Pi _{\mu \nu }^{reg}(k)\left| _{k=0}\right. =0$, \qquad
\qquad (10.7)

must be physically interpreted as the conditions for the completeness of
physical photon formation and for the impossibility of its self-acceleration.

Thus it is stated that the subtraction regularization corresponds to
mathematical formulation of the common physical conditions, primary imposed
on the system, and therefore it is far from an artificial, ad hoc method.

It must be specially underlined that the method of renormalization group
[37, 31] can be immediately interpreted via the temporal functions. Really,
as the corresponding Lie equations contain logarithmic derivatives of
propagators over energy-moment, they are still proportional to the temporal
magnitudes.

As the denominators of propagators leads only to the trivial terms,
connected with the uncertainty principles or twice bigger them, let's
consider for checking of such {\it proposition} the nondimensional Green
functions $\breve{G}(q)$ with all 4-moments, except one, fixed. Then in
accordance with the renorm-group equation of Callan and Symanzik [38] it can
be written that

$|\xi _{\mu }|\ \symbol{126}\ q^{2}\frac{\partial }{\partial q^{2}}\ln
\breve{G}(q)=(\gamma _{m}-1)\frac{\partial }{\partial m^{2}}\ln \breve{G}%
+\beta \frac{\partial }{\partial e}\ln \breve{G}-\gamma _{G}(m^{2},e)$,
\qquad \qquad (10.8)

where $\gamma _{m}$, $\beta $ and $\gamma _{G}$ are the structure functions
of the renorm-group.

In the lowest order of $\varphi ^{4}$ theory $\gamma _{m}=0$, $\beta =\frac{3%
}{2}e^{2}$ and $\gamma _{G}=-\frac{3}{2}e$. Therefore in (10.8) for the
4-tail graphs are retained only the terms connected with the charge
formation and accumulation of the observed mass:

$e\frac{\partial }{\partial e}\ln \breve{G}=e(1-\breve{G}^{-1})$;

$-\frac{\partial }{\partial m^{2}}\ln \breve{G}=-\frac{e^{2}}{m^{2}\breve{G}}%
\mathop{\displaystyle\sum}%
\frac{1}{y_{k}}\ $Arth $y_{k}$, \qquad \qquad (10.9)

where $y_{k}=(z_{k}^{2}-z_{k})^{1/2}$, $(z_{1},z_{2},z_{3})=(s,u,t)/m^{2}$.

It shows that in the $\varphi ^{4}$ theory, and correspondingly in the QED,
the charge increasing must extend the duration of formation, but in such
gauge theories, where $\beta <0$, this process should decrease $\xi _{\mu }$.

Note that in the UV limit $\gamma _{m}=1$ in (10.8) and

$\ln \breve{G}(q^{2},e)\rightarrow \ \ln \breve{G}(1,e)-2\gamma _{G}\ln q$,
\qquad \qquad (10.10)

Hence in the asymptotically free theories, where $e\rightarrow -e$, the
expression (10.8) can be reduced to such relation:

$\xi _{\mu 2}(k)=\frac{2q_{\mu }}{q^{2}}(1-6\frac{|e|}{\nu })$, \qquad
\qquad (10.11)

i.e. at $|e|=1/6\nu $ the duration of formation in this approximation is
equal to zero.

This result can be of general interest: it seems very tempting to attempt
this form with problem of existence of only restrict types of particles
generation, but this invites further more detailed investigations. Note that
the absence of terms, which describe the delays in the (10.8), can be
connected with the calculation of matrix elements in the one-loop
approximation.

\qquad

{\large CONCLUSIONS}

The main results of performed researches can be formulated in such points.

1.There are established the reciprocal forms of the Schr\"{o}dinger or the
Bloch equations in $p$-space and their covariant generalizations. By these
equations are substantiated some different, as has been seemed, methods of
calculation of durations of scattering.

2.The reciprocal equations permit, in particular, the unified determination
of the Wigner-Smith function of delay at scattering process and the
Pollak-Smith function of duration of final state formation.

3.The magnitudes of duration of scattering process (time of delay and
duration of formation of final state) are implicitly contained in the usual
field theory. The propagators of interacting fields actually can describe
them and thereby it becomes evident why many problems of kinetics could be
considered without explicit introducing of temporal magnitudes. Just it may
explain a delayed beginning of researches of temporal parameters in the
quantum theory.

4.The adiabatic hypothesis of quantum theory may be considered as the
implicit expression of existence of the certain duration of formation
(dressing) of physical particles. Therefore it is far from the formal, pure
mathematical procedure and shows that at the consideration of arising of any
physical state the account of its formation duration is inevitable.

5. Both basic methods of delay duration measurement, by Wigner and Smith and
by ''Larmor clock'', can be described as an addition of zero-energy line to
the Feynman graph of process.

6. The dispersion relations for temporal functions are established. They
prove, in particular, that the duration of final state formation is, at
least, twice bigger than the uncertainty values and therefore is measurable.

7. The consideration of the Lorentz (oscillator) model of simple dispersing
medium leads to an intuitively evident interpretation of temporal functions.
In this model the function of time delay is proportional to the polarization
of medium and the function of final state formation is proportional to the
electric conductivity.

8. The transition from the Schr\"{o}dinger equation to the reciprocal
temporal equation corresponds to the Legendre transformation of classical
action function. The averaged values of these functions are deduced via the
Fourier transformation of response functions. The covariant form of temporal
equation is deduced by the temporal variant of Stueckelberg-Bogoliubov
variational method.

9. The methods of subtraction regularization in field theory can be
physically substantiated and explained as the imposing the requirements of
finishing of their formation and impossibility of particles
self-acceleration on propagators of particles.

10. The concept of duration of interactions imparts the evident physical
sense to the equations of renormalization group and demonstrates that the
formation of each of particles parameters required of the certain (may be,
specific) duration. Hence it gives a possibility to think that the
coordination of durations of these partial processes will allow more
detailed understanding of peculiarities of those or other particles. Such
program requires, however, further researches.

We do not discuss here a lot of delays determinations known in the current
literature. It can be suggested that just the revealed analytical properties
of the composed temporal functions demonstrate their general significance.
It does not exclude, of course, the possible usefulness of some other
determinations in special cases.

Note that apart from the general problems of interpretation, the represented
approach simplify the consideration of such phenomena as multiphoton
processes, in which the termination of process depends on intervals between
subsequent interactions [6, 15], the general theory of optical dispersion
[29]; it permits to consider the superluminal phenomena as processes
connected with the properties of near fields of radiation [22], etc. Some
other applications of this approach will be considered elsewhere.

\qquad

\bigskip \bigskip {\large ACKNOWLEDGMENTS}

I am deeply indebted to discussions with A.D.Sakharov on the early stages of
the investigations. A lot of advantage the author has received from
discussions of these subjects per different years with M.Ya.Amusia,
R.Englman, V.Ya.Fainberg, G.Mainfray, E.Pollak, I.I.Royzen, G.M.Rubinstein.

\qquad

{\bf REFERENCES}

[1]. L.A.Zadeh, Ch.A.Desoer. {\it Linear System Theory. (The State Space
Approach)}. NY: McGraw-Hill, 1967; E.Skudrzyk. {\it The Foundations of
Acoustics (Basic Mathematics and Basic Acoustics)}. NY: Springer, 1971.

[2]. D.Bohm. {\it Quantum theory}. Prentice-Hall, NY: 1952, Ch.11.

[3]. E.P.Wigner. Phys.Rev., {\bf 98}, 145 (1955), Wigner cited in this
connection the unpublished theses of L.Eisenbud.

[4]. F.T.Smith. Phys.Rev., {\bf 118}, 349; {\bf 119}, 2098 (1960); {\bf 130}%
, 394 (1963); J.Chem.Phys., {\bf 36}, 248;{\bf \ 38}, 1304 (1963).

[5]. M.L.Goldberger, K.M.Watson. {\it Collision Theory}. Wiley, NY, 1964.

[6]. M.E.Perel'man. Phys.Lett. {\bf 32}A, 64 (1970); Sov.Phys. JETP, {\bf 31}%
, 1155 (1970) [Zh.Eksp.Teor.Fiz., {\bf 58}, 2139 (1970)]; M.E.Perel'man,
V.G.Arutyunian, Sov.Phys. JETP, {\bf 35}, 260 (1972) [Zh.Eksp.Teor.Fiz.,
{\bf 62}, 490 (1972)].

[7]. M.E.Perel'man. Sov.Phys. Doklady, {\bf 19}, 26 (1974) [DAN SSSR, {\bf %
214}, 539 (1974)].

[8]. A.I.Baz'. Sov.J.Nucl.Phys., {\bf 4}, 229 (1967).

[9]. E.H.Hauge, J.A.St{\small \O }vneng. Rev.Mod.Phys. {\bf 61}, 917 (1989);
R.Landauer, Th.Martin. Rev.Mod.Phys. {\bf 66}, 217 (1994); J.G.Muga,
C.R.Leavens. Phys.Rep., {\bf 338}, 353 (2000) and references therein.

[10]. {\it Time in Quantum Mechanics} (J.G.Muga e.a., Ed's). Springer, 2002.

[11]. I.M.Frank. Izvestia Akademii Nauk SSSR, ser. fizich., {\bf 6}, 10
(1942).

[12]. M.L.Ter-Mikaelyan. Zh.Eksp.Teor.Fiz., {\bf 25}, 289 (1953).

[13]. L.D.Landau, I.Ya.Pomeranchuk. DAN SSSR, {\bf 9}2, 535, 735 (1953).

[14]. M.L.Ter-Mikaelyan. {\it High Energy Electromagnetic Phenomena in Medium%
}. NY, Wiley, 1972; E.L.Feinberg. UFN, {\bf 132}, 255 (1980)
[Sov.Phys.Uspekhi, {\bf 20}, 629 (1980)]; B.M.Bolotovskii. Proc.Lebedev
Phys.Inst. {\bf 140}, 95 (1982) and references therein.

[15]. M.Moshinsky. Phys.Rev., {\bf 81}, 347 (1951); {\bf 84}, 525 (1951).
See, also, the interpretation of this phenomenon: V.Man'ko, M.Moshinsky,
A.Sharma. Phys.Rev.A {\bf 59}, 1809 (1999)

[16]. M.E.Perel'man. {\it Kinetical Quantum Theory of Optical Dispersion}.
Tbilisi, Mezniereba, 1989 (In Russian); in: {\it Multiphoton Processes}
(G.Mainfray \& P.Agostini, Ed's) Paris, CEA, 1991, 155-165.

[17]. E.Pollak, W.H.Miller. Phys.Rev.Lett., {\bf 53}, 115 (1984); E.Pollak.
J.Chem.Phys., {\bf 83}, 1111 (1986).

[18]. J.L.Synge. {\it Classical Dynamics}. In: {\it Handbuch d. Physik}
(Hrsg. S.Flugge). {\bf III/1}, Springer, 1960; V.A.Arnold. {\it Mathematical
Methods of Classical Mechanics}. Moscow, 1974.

[19]. M.E.Perel'man. Zh.Eksp.Teor.Fiz., {\bf 50}, 613 (1966) [Sov. Phys.
JETP, {\bf 23}, 487(1966)]; Dokl. Ak. Nauk SSSR, {\bf 187}, 781 (1969)
[Sov.Phys.Doklady, {\bf 14},772 (1969)]; Bull. Acad. Sc. Georgian SSR, {\bf %
81}, 325 (1976).

[20]. M.E.Perel'man, R.Englman. Mod.Phys.Lett., {\bf 14}, 907 (2000).

[21]. W.Pauli. {\it Quantentheorie}. In: {\it Handbuch d. Physik} (Hrsg.
H.Geiger, K.Scheele). Bd. {\bf 23}, Springer, 1923.

[22]. M.E.Perel'man. In: arXiv. physics, Gen.Phys/0309123 (30/09/2003)

[23]. E. Schr\"{o}dinger. Sitzugsber.Preuss.Akad.Wiss., 1930, p. 296.

[24]. E.P.Wigner. In: {\it Aspects of Quantum Theory}. (A.Salam, E.P.Wigner,
Ed's). Cambridge, 1972, p.237.

[25]. M.W.Mitchell, R.Y.Chiao. Am.J.Phys., {\bf 66}, 14 (1998); T.Nakanishi,
K.Sugiyama, M.Kitano. Am.J.Phys., {\bf 70}, 1117 (2002). See also:
R.Y.Chiao, J.M.Hickmann, C.Ropers, D.Solli. In: N. Bigelow (Org.). {\it %
Coherence and Quantum Optics}. {\bf VIII}. 2002.

[26]. V.L.Ginzburg. {\it The Propagation of Electromagnetic Waves in Plasmas}%
. Pergamon, 1970.

[27]. L.D.Landau, E.M.Lifshitz. {\it Electrodynamics of continuous media}.
L.: Pergamon (any edition).

[28]. M.Hillery, R.F.O'Connell, M.O.Scully, E.P.Wigner. Phys.Rep., {\bf 106}%
, 121 (1984) and references herein.

[29]. S.R.de Groot, W.A.van Leeuwen, Ch.G.van Weert. {\it Relativistic
Kinetic Theory (Principles and Applications)}. North-Holland, Amsterdam,
1980 and references herein..

[30]. M.E.Perel'man, G.M.Rubinstein. Sov.Phys. Doklady, {\bf 17}, 352 (1972)
[DAN SSSR, 203, 798 (1972)].

[31]. E.Rosenthal, B.Segev. Phys.Rev.A, {\bf 66}, 05210 (2002).

[32]. E.C.G.Stueckelberg. Phys.Rev. {\bf 81}, 130 (1951).

[33]. N.N.Bogoliubov, D.V.Shirkov. {\it Introduction to the Theory of
Quantized Fields}. 3rd Ed.,Wiley, 1980.

[34]. T.D.Newton, E.P.Wigner. Rev.Mod.Phys., {\bf 21}, 400 (1949).

[35]. M.Hawton. Phys.Rev.A, {\bf 59}, 954 (1999) and references herein..

[36]. V.Berestetski, E.Lifshitz, L.Pitayevski. {\it Relativistic Quantum
Theory}. Pergamon, Oxford, 1971.

[37]. E.C.G.Stueckelberg, A.Peterman. Helv.Phys.Acta., {\bf 24}, 153 (1953);
M.Gell-Mann, F.Low. Phys.Rev., {\bf 95}, 1300 (1954).

[38]. C.G.Callan. Phys.Rev.D, {\bf 2}, 1541 (1970); K.Symanzik.
Comm.Math.Phys., {\bf 18}, 227 (1970). The general review in: C.Itykson,
J.-B.Zuber. {\it Quantum Field Theory}. McGraw, NY, 1980.

\end{document}